\documentclass[aps,prl,twocolumn,showpacs,superscriptaddress,groupedaddress]{revtex4}

\usepackage{graphicx} 
\usepackage{dcolumn} 
\usepackage{bm}
\usepackage{amssymb}
\usepackage{comment}
\begin{document}

\widetext
\title{Molecular Orientation and Surface Site Dependence of the Dissociative Adsorption of O$_{2}$/Al(111).}

\affiliation{Department of Applied Physics, Osaka University, Suita, Osaka 565-0871, Japan}
\author{K.~Shimizu} \affiliation{Department of Applied Physics, Osaka University, Suita, Osaka 565-0871, Japan}
\author{W.A.~Di\~{n}o} \affiliation{Department of Applied Physics, Osaka University, Suita, Osaka 565-0871, Japan}
\author{H.~Kasai} \affiliation{Department of Applied Physics, Osaka University, Suita, Osaka 565-0871, Japan}
\noaffiliation
\vskip 0.25cm

\date{\today}

\begin{abstract}
We reproduced the initial sticking probability of O$_{2}$/Al(111) by use of spin-polarized density functional theory and quantum dynamics calculations.
We found a large activation barrier when the molecule is dissociatively adsorbed through $abstraction$ on the top site unlike on the hollow and bridge sites.
This barrier is sensitive to the molecular orientation and surface site.
The vibrational motion of the molecule facilitates the abstraction due to the $late~barrier$.
Finally, we suggest a possibility on how widely separated oxygen atoms adsorbed on the surface could be realized.
\end{abstract}

\maketitle

Oxidation is one of the most common chemical reaction, and generally it happens by O$_{2}$ molecules in the atmosphere.
In addition, oxides are used for wide variety of industrial applications, e.g., fuel cell and photocatalyst.
As just described, oxygen plays a critically important role in this world.
For metal surfaces in a microscopic point of view, the dissociative adsorption of the O$_{2}$ molecule is considered as a first step of the oxidation.
Afterward the adsorbates diffuse onto the surface and/or penetrate into the subsurface by excess energy to form an oxide.
From these importance, a lot of studies have been conducted for oxygen reactions.

Aluminum is often taken as a prominent material to a benchmark of oxidation due to its simple electronic structure and wide range of applications.
For this surface, however, even the first step of oxidation has not yet been resolved.
It is found from molecular beam experiment that the initial sticking probability of O$_{2}$ molecule on Al(111) surface, including chemisorption and physisorption, is low at low initial translational energy~[\onlinecite{Kasemo}]; it simply suggests an existence of activation barrier.
On the other hand in the theoretical studies, it is indicated that there is no activation barrier for dissociative adsorption~[\onlinecite{Lundqvist},\onlinecite{Ohno}].
This discrepancy between experiment and theoretical studies has been a long standing problem.
Moreover, it is also puzzling that the scanning tunneling microscopy (STM) image shows a widely separated adsorbates ($\sim$ 80 \AA) at low oxygen surface coverage~[\onlinecite{Brune}].

In the present Letter, we investigated six-dimensional potential energy surfaces (PESs), and it is shown that the PESs are of strikingly difference by orientation of O$_{2}$ molecule and surface adsorption sites.
After that, from the perspective of dynamics, we reproduced the first initial sticking probability under adiabatic picture qualitatively in agreement with the experimental result.
Finally, we will offer a possibility of how the adsorbates are separated such a long distance.

We constructed the PESs by total energy calculations based on spin-polarized density functional theory (DFT)~[\onlinecite{VASP1},\onlinecite{VASP2}] by varying all degrees of freedom of the molecule, using the projector augmented wave (PAW) method to describe the electron-ion interactions and the generalized gradient approximation (GGA), within Perdew-Burke-Ernzerhof (PBE), for the exchange-correlation energy~[\onlinecite{PBE}].
The Kohn-Sham equations are solved using plane waves with kinetic energy up to 800 eV.
Brillouin zone integration is performed using the special-point sampling technique of Monkhorst and Pack, 5$\times$5$\times$1 sampling meshes~[\onlinecite{Monk}].
The slab model used here consists of seven Al layer with each substrate layer containing four Al atoms, and each slab is separated by ca. 20 \AA~vacuum.
We used the unrelaxed surface for the whole calculations since our aim is to get the initial sticking probability.
Although a constrained DFT calculation is developed to restrict the spin state of O$_{2}$ molecule for dissociation~[\onlinecite{Scheffler},\onlinecite{Gross}], i.e., transition from spin-triplet to singlet is forbidden, we adopted the conventional spin-polarized DFT because such spin states might be treated correctly.

In order for the molecule to dissociate and adsorb on the surface, we must define a path of least potential, that is called reaction path along which the reaction proceeds, from PESs.
We did quantum dynamics calculations of the sticking probability by solving the time-independent Schr\"odinger equation for an O$_{2}$ molecule moving under the influence of PESs corresponding to the O$_{2}$/Al(111) system using the coupled-channel method~[\onlinecite{Kasai},\onlinecite{10}] and the concept of a local reflection matrix and an inverse local transmission matrix~[\onlinecite{Brenig},\onlinecite{Brenig2}].
The degrees of freedom we considered are the center-of-mass (CM) distance of the molecule from the surface, $z$, interatomic distance of the molecule, $r$, and the surface coordinate, $x$, respectively.
For the surface coordinate $x$, we chose the $\langle$11$\bar{2}$$\rangle$ direction.

In this study, we assumed that the dissociative adsorption happens through $abstraction$, in which bottom oxygen atom adsorbs on the surface and upper one moves away from the surface when the O$_{2}$ orientation is non-parallel to the surface.
Hence, for the dynamics calculations we fixed the polar orientation of the molecule perpendicular to the surface, and froze the rotational motion.
In fact, a large number of channels are needed for rotational motion in the O$_{2}$ molecule due to its relatively large mass.
Higher multidimensional calculations than conducted in this Letter exceed the limitation of our calculation memory.
This assumption is supported by the experiment in which the atomic oxygen is observed during the molecular beam experiment by use of Resonance Enhanced Multi-Photon Ionization (REMPI)~[\onlinecite{Binetti}].
Recently, experimental techniques are developed so that the rotational quantum state of the O$_{2}$ molecule can be manipulated in the molecular beam apparatus by using an electrostatic hexapole~[\onlinecite{Okada},\onlinecite{Kurahashi}].
Thus, we can now specifically extract the information which we want to know in reactions, e.g., reactions of a particular orientation or surface site.
This seems to be compatible with the implementation of quantum dynamics calculations.

\begin{figure}
\includegraphics[scale=0.4]{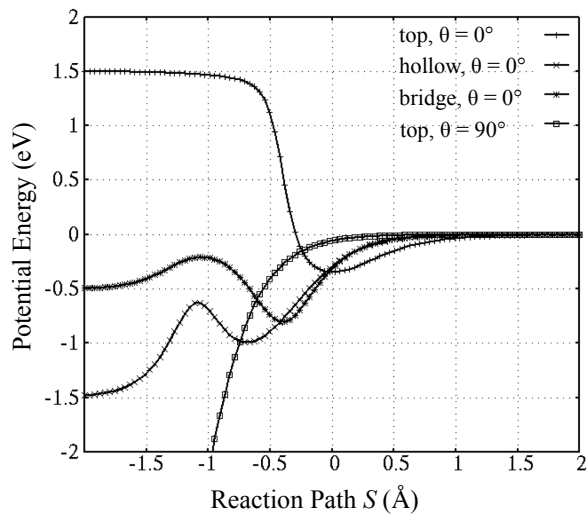}
\caption{\label{pec} PECs of O$_{2}$ dissociation on the Al(111) surface as a function of the reaction path $S$. The positive side of reaction path is gas phase, and negative is on the surface. Adsorption sites are top, beirge, and hollow, and polar orientations are perpendicular or parallel with respect to the surface.}
\end{figure}

As already shown in the previous study~[\onlinecite{Lundqvist}], we also found that there is no activation barrier for dissociative adsorption of O$_{2}$ molecule on Al(111) surface when the molecule approaches the surface in a parallel polar orientation with respect to the surface.
It is also in accord with the previous study that there is one exception where a slight barrier exists when the CM of the molecule is on the bridge site and the molecular axis is directed to the hollow site.
This tendency of no barrier is the same regardless of which surface site the molecule impinges on.
In each configuration, the molecule obtained the considerably large adsorption energy, especially adsorption on the hollow site where this energy is ca. 6.6 eV.
With the molecule coming closer to the surface, the hybridization between $p$-orbital of oxygen atom and $s$-orbital of aluminum atom is formed, and this can be seen in the local density of state~[\onlinecite{Shimizu}].
Because oxygen has strong electronegativity and aluminum has relatively small ionization energy, the electrons of the surface transfer to the molecule readily.
Since such electrons fill the oxygen $\pi^{\ast}$ anti-bonding orbitals, the intermolecular bond becomes weak, elongates, and finally breaks without barrier.
It could be said that after chemically adsorbed, the magnetization of oxygen derived from unpaired electron disappears due to such electron transfer. 

For the case of perpendicular polar orientation with respect to the surface, where dissociation occurs through abstraction, behaviors of the O$_{2}$ molecule are quite different depending on the adsorption sites.
Figure 1 shows the potential energy curves (PECs) along the reaction path for abstraction on the bridge, hollow, and top sites.
As can be seen from PECs, there is no activation barrier for on the bridge and hollow sites.
We can therefore justify the occurrence of spontaneous abstraction.
On the other hand, we unexpectedly found a large activation barrier (ca. 1.5 eV) on the top site.
Because this barrier is quite large, we investigated low symmetry sites and found that the surface area of the activation sites, where there is an activation barrier, is the majority due in part to the highest atomic density of face-centered cubic (fcc) (111) facet.
In the case of no barrier sites, the orbitals of O$_{2}$ molecule can overlap with the sea of electrons of the aluminum surface when the molecule comes close to the surface.
Hence, it is easy to form hybridization, and the abstraction happens easily.
On the other hand, because such sea is distributed parallel to the surface and charge density is low on the top site, the hybridization seldom occurs when the molecule approach the top site.
Thus, the molecule needs the large energy for abstraction in this case.
The Al(111) surface can not be considered as flat unlike the case of H$_{2}$/Cu(111)~[\onlinecite{WAD}].

Whether an activation barrier exists or not is mainly due to the position where the bottom oxygen atom impinges.
Thus, for the case of tilted polar orientations with respect to the surface, 30$^{\circ}$, 45$^{\circ}$, and 60$^{\circ}$, when the bottom oxygen atom approached around the top site, we found that an activation barrier exists as well~[\onlinecite{Shimizu}].

\begin{figure}[b]
\includegraphics[scale=0.45]{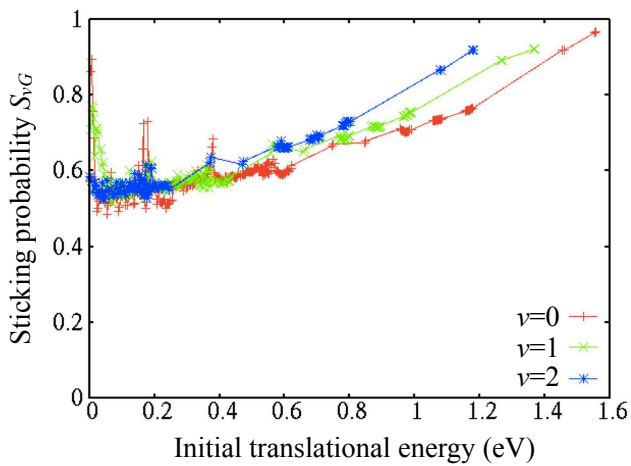}
\caption{\label{SvG} Sticking probability $S_{vG}$ as a function of the initial translational energy of the O$_{2}$ molecule. The probabilities are separated by the initial vibrational state $v$ = 0, 1, and 2, respectively. The motion parallel to the surface are summed by final state. }
\end{figure}

The adsorption energy is considerably larger for the parallel polar orientation than the perpendicular one as shown in Fig.~\ref{pec}.
Comparing their CM motions on the top site, however, the potential energy is lower for the perpendicular orientation case due to the physisorption energy well at reaction path $S$~$\sim$~0.
Thus, the O$_{2}$ molecule could be assumed to steer to the perpendicular orientation at thermal translational energy range.
On top of this, because the molecular beam experiment is based on the King-Wells (KW) method~[\onlinecite{King}], in which the initial sticking probability is calibrated by the change of the oxygen partial pressure, the molecule randomly impinges on the surface.
Furthermore, the PESs are quite different depending on the adsorption sites.
Thus, we took the surface corrugation into account for dynamics calculations.

Figure~\ref{SvG} shows the initial sticking probability of O$_{2}$ on Al(111) as a function of the initial translational energy of the molecule, and the incident angle is normal to the surface.
The curves are separated by the initial vibrational state of the molecule, $v$ = 0, 1, and 2, respectively.
When the initial translational energy is around zero, the sticking probability is high, e.g., ca. 0.9 for the vibrational ground state.
This is caused by the steering of O$_{2}$ along surface, i.e., the molecule initially impinging on the top site moves to the potential well on the bridge or hollow sites.
The sticking probability decreased immediately with the increase of the translational energy until ca. 0.5 because the molecule was reflected by the barrier before evolution of the steering.
This value is almost the same with the surface area of the activation site.
Afterward the probability increased with increasing the translational energy.
This is qualitatively in agreement with experimental result.

At higher translational energy the calculated sticking probability was smaller than the experimental result.
However, the probability would be enhanced for the dynamics calculations including the parallel orientation at higher translational energy because the parallel orientation is the open channel for dissociation.
It would be necessary to take into account the rotational motion for more precise simulations.
Besides, the experimental sticking probability does not reach unity in the whole initial translational energy range.
This might be due in part to the abstraction as depicted from the KW method, i.e., some of the oxygen atom would remain in gas phase and influence the partial pressure.

An oscillatory structure appeared at small initial translational energy range in Fig.~\ref{SvG} resembling the case for H$_{2}$/Pd(100)~[\onlinecite{Gross2}].
This phenomenon is due to the motion parallel to the surface of incident molecule.
With increasing the translational energy, the new channel for the motion parallel to the surface opens so that the sticking probability increases.
This is the typical phenomenon for the system which includes non-activated as well as activated path.

\begin{figure}
\includegraphics[scale=0.45]{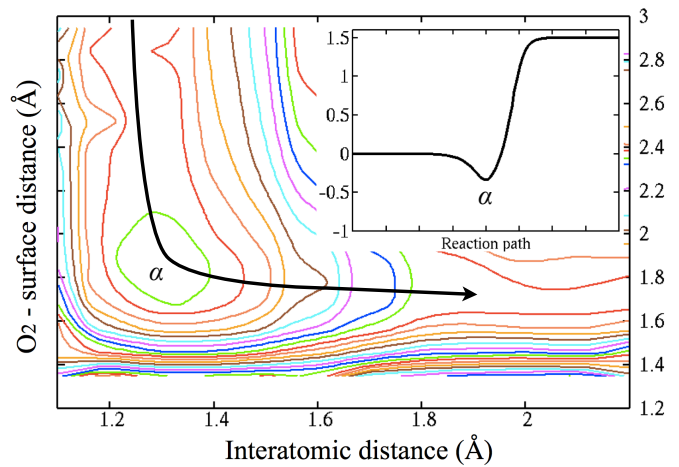}
\caption{\label{pes}  PES of O$_{2}$ molecule on the Al(111) surface with interatomic distance and a distance between bottom oxygen atom and surface. The polar orientation is perpendicular with respect to the surface and the adsorption site is fcc-hollow site. The contour interval is 0.2 eV. The arrow indicates the reaction path which is the path of least potential. The point $\alpha$ is corresponding to the physisorption site and curved region. The inset shows the PEC in this case.}
\end{figure}

We can see that the sticking probability became higher for higher initial vibrational state of the molecule in accordance with the experiment.
This shows the evolution of the vibrational assisted sticking (VAS) effect, which is already familiar to H$_{2}$ molecule~[\onlinecite{Miura}].
As shown in PES in Fig.~\ref{pes} for the top site as an example, the reaction path curves before the molecule encounteres the activation barrier, this is called $late~barrier$~[\onlinecite{Holloway}].
The vibrational energy of the molecule $\hbar\omega$ is initially ca. 189 meV in gas phase.
But after dissociation the vibration is between the adsorbate and substrate, and it is smaller than that in gas phase, ca. 50 meV.
This decrement is lager for the higher vibrational state, and it occurs along the curvature of the reaction path around Point $\alpha$ in Fig.~\ref{pes}, where the translational motion and vibrational motion are coupled.
The decreased energy transfers to the translational energy by the law of conservation of energy.
Hence, the effective translational energy helps the molecule overcome the barrier.
An ambiguity for VAS effect indicated in the experiment~[\onlinecite{Kasemo}] is eliminated in here.

Until now, this surface sensitivity has been hindered by the fixed concept that the O$_{2}$ molecule impinges only on the hollow site.
We reproduced the sticking probability qualitatively in agreement with the experiment by the consideration of the surface corrugation.
In addition, we found the tunneling effect of the O$_{2}$ molecule although it might not be so large to affect the reaction.

\begin{figure}
\includegraphics[scale=0.45]{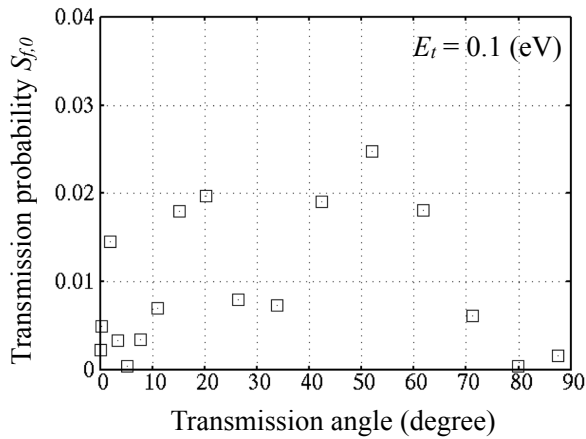}
\caption{\label{reflection} The transmission probability of the molecule as a function of the transmitted angle. The incident angle of the molecule is initially normal to the surface at the initial translational energy $E_{t}$ is 0.10 eV. }
\end{figure}

We show the state resolved transmission probability in Fig.~\ref{reflection}, where the initial translational energy was 0.10 eV and the sticking probability was ca. 0.50.
The incident angle is normal to the surface, and final state $f$ is projected onto the transmission angle.
The transmitted molecule spreads to the wide angle, and the most part of the molecule changes its angle from normal.
In light of the abstraction, the upper oxygen atom may move away from the surface at an angle when the transmitted molecule change its motion parallel to the surface.
This atom interacts with the surface, and it finally adsorbs on the surface because it is more stable to adsorb on the surface than to stay as an atom in the gas phase~[\onlinecite{Shimizu}].
Moreover, because the position where this atom starts to move away is relatively far from the surface, there is no barrier for the migration on the surface.
Therefore, the migration of the atom moving away from the surface at an angle might have a possibility of the widely separated adsorbates on the Al(111) surface as mentioned earlier.

In summary, we reproduced the initial sticking probability of the O$_{2}$/Al(111), owing to the large PESs data in which we found the large activation barrier for abstraction on the top site unlike on the hollow and bridge sites.
The vibrational motion facilitates the dissociative adsorption even when the reaction happens through abstraction.
At the end, we suggested the mechanism that the oxygen atom which has the possibility to move away from the surface at an angle would lead to the large distance between oxygen atoms adsorbed on the surface.
We conclude that the reaction of the molecule must be considered from the view of the dynamics.
With the advancement of experimental devices, soon it will be possible to calibrate experimentally the sticking probability of fixed orientation along with the desired surface site.

This work is supported in part: by ISSP (University of Tokyo), KEK (No.T10-12), Cybermedia Center (Osaka University), YITP (Kyoto University), NIFS, MEXT (Ministry of Education, Culture, Sports, Science and Technology) through the G-COE (Special Coordination Funds for the Global Center of Excellence) program "Atomically Controlled Fabrication Technology", Grant-in-Aid for Scientific Research on Innovative Areas Program (2203-22104008) and Scientific Research (c) (22510107) program; by JST (Japan Science and Technology Agency) through ALCA (Advanced Low Carbon Technology Research and Development) Program.

\begin{comment}
\end{comment}

\end{document}